\documentclass[prl,twocolumn,showpacs,superscriptaddress]{revtex4}

\usepackage{epsfig}
\usepackage{amssymb}
\usepackage{graphicx, natbib}

\begin{document}

\title{QND measurement of a superconducting qubit in the weakly projective regime}

\author{T. Picot}
\affiliation{Kavli Institute of Nanoscience, Delft University of Technology, PO Box 5046, 2600GA Delft, The Netherlands}
\author{R. Schouten}
\affiliation{Kavli Institute of Nanoscience, Delft University of Technology, PO Box 5046, 2600GA Delft, The Netherlands}
\author{C.J.P.M. Harmans}
\affiliation{Kavli Institute of Nanoscience, Delft University of Technology, PO Box 5046, 2600GA Delft, The Netherlands}
\author{J.E. Mooij}
\affiliation{Kavli Institute of Nanoscience, Delft University of Technology, PO Box 5046, 2600GA Delft, The Netherlands}

\date{ \today}

\begin{abstract}

Quantum state detectors based on switching of hysteretic Josephson junctions biased close to their critical current are simple to use but have strong back-action. We show that the back-action of a DC-switching detector can be considerably reduced by limiting the switching voltage and using a fast cryogenic amplifier, such that a single readout can be completed within 25 ns at a repetition rate of 1 MHz without loss of contrast. Based on a sequence of two successive readouts we show that the measurement has a clear quantum non-demolition character, with a QND fidelity of 75 \%.

\end{abstract}

 \maketitle

According to the Heisenberg uncertainty principle, the measurement of a physical variable necessarily perturbs its quantum conjugate variable, imposing a fundamental limit on the precision of the measurement. However, it is in principle possible to decouple the dynamics of the variable which is measured from the perturbed conjugate variable. This is the central idea in the concept of Quantum Non-Demolition (QND) measurement, which was initially developed to perform high precision measurement beyond the standard quantum limit \cite{Braginski}. QND measurement became an important paradigm of quantum mechanics of particular interest for fundamental studies of quantum systems. A simple criterion ensuring that a measurement is QND is that the Hamiltonians of the bare quantum system and of the interaction between the quantum system and the detector do commute. In practice, its realization is not straightforward as it requires a very good control of the back-action of the classical detector on the state of the quantum system. So far, QND measurements have been demonstrated experimentally in quantum optics \cite{Grangier}, in cavity-QED with Rydberg atoms \cite{Guerlin}, with trapped single electrons \cite{Peil}, and more recently with superconducting circuits \cite{Lupascu,Boulant,Johnson} based on dispersive coupling of a quantum 2-level system (or qubit) with a resonator-type detector.

In this letter we study the quantum measurement of a superconducting flux qubit, employing a hysteretic dc SQUID, which is either in a superconducting state or in a dissipative state. The back-action of this detector, in particular if in the dissipative state, is commonly much higher than the minimum imposed by quantum mechanics, thus seriously limiting its efficiency. In our experiment, we show that a single readout can be completed in a time as short as 25 ns, and that the back-action can be controllably reduced to such a level that for the first time with this type of detector, a QND measurement is possible with an averaged QND fidelity of 75 \%. It should be stressed that although this type of fast DC-switching detector can generally provide high readout contrast\cite{Claudon}, in this experiment we chose a small qubit-detector interaction resulting in a low readout contrast and therefore a weakly projective measurement. This simple and fast QND detector forms a highly attractive alternative for dispersive QND detectors.

\begin{figure}[t]
\includegraphics[width=3.4in]{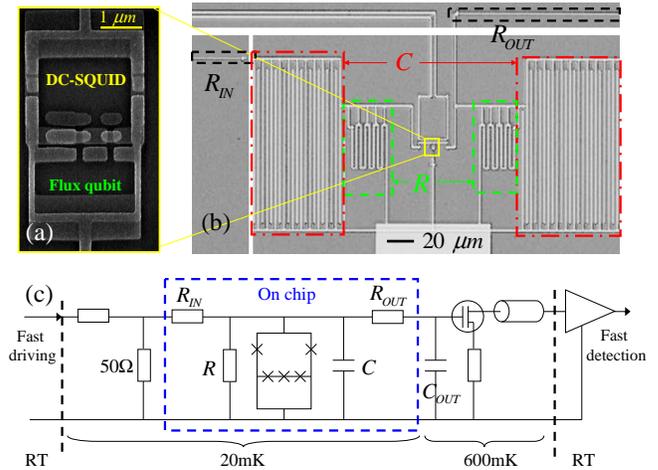}\caption{ \label{sample} (color on-line) (a) Scanning electron microscopy picture of the flux qubit and the dc SQUID. (b) Optical picture of the readout circuit. (c) Schematic of the readout circuit including a dc SQUID shunted with a resistor $R=125$ $\Omega$ and a capacitor $C=1$ pF, and a HEMT cryogenic amplifier allowing a very fast detection of the qubit state. }
\end{figure}

The measurement can be simply illustrated in the Bloch sphere representation of the qubit state. Denoting as $|g>$ and $|e>$ the ground state and excited state respectively, the general qubit state can be written as $|\psi>=\cos{\left(\frac{\theta}{2}\right)}|g>+e^{i\phi}\sin{\left(\frac{\theta}{2}\right)}|e>$. The variable which is measured is the qubit excited state occupation probability $p(e)=\sin^2{\frac{\theta}{2}}$, while $\phi$ is the perturbed conjugate variable. As the free evolution of $\theta$ is decoupled from $\phi$, the main requirement for a QND measurement is that no transitions between $|g>$ and $|e>$ occur during the measurement. Due to the small discrimination efficiency between $|e>$ and $|g>$ in a single readout, $p(e)$ is obtained from the averaged outcome of a large ensemble of identical qubit state preparation and readout sequences. In the case of an ideal QND detector, the averaged outcome of two successive measurements should be equal. Additionally, the correlation of the measurement outcomes reveals the degree of projectiveness of the measurement which in our experiment is very small.

The flux qubit, shown in Fig. \ref{sample}a, is made of a superconducting loop interrupted by four Josephson junctions (the largest junction being auxiliary), biased with an external flux $\Phi$ close to half a flux quantum $\frac{1}{2}\Phi_0 = h/4e$ \cite{Hans}. In general, the two energy eigenstates $|g>$ and $|e>$ are quantum superpositions of two states corresponding to oppositely circulating persistent currents of magnitude $I_p$ and denoted as $|+>$ and $|->$. In the basis of these current states the Hamiltonian of the flux qubit can be written as: $\hat{H}=-\frac{1}{2}(\epsilon \hat{\sigma_z} + \Delta \hat{\sigma_x})$, where $\hat{\sigma_x}$, $\hat{\sigma_z}$ are Pauli spin matrices, $\epsilon=2 I_p (\Phi-\Phi_0/2)$ and $\Delta$ is the tunnel coupling between the two current states. The difference of energy between $|g>$ and $|e>$ is $E = E_e-E_g = \sqrt{\epsilon^2+\Delta^2}$. All the measurements are performed at a bias $|\epsilon|>>\Delta$, where the energy eigenstates $|g>$ and $|e>$ approximately coincide with the current states $|+>$ and $|->$.

The quantum state of the qubit is read out by measuring the flux resulting from the expectation value of the circulating current, using a hysteretic dc SQUID comprising two Josephson junctions in a superconducting loop. The interaction between the qubit and the detector is given by $\hat{H_I} = M I_p \hat \sigma_z J$, where $M$ is the qubit-dc SQUID mutual inductance and $J$ is the circulating current in the SQUID loop. We should stress that the mentioned similarity between the eigenstates of the current and the energy for $|\epsilon|>>\Delta$ closely approaches the sufficient condition for a QND measurement $[\hat{H},\hat{H}_I]=0$. Furthermore, the large difference between the internal frequency of the detector and the qubit frequency suppresses energy exchange between qubit and detector.

\begin{figure}[t]
\includegraphics[width=3.4in]{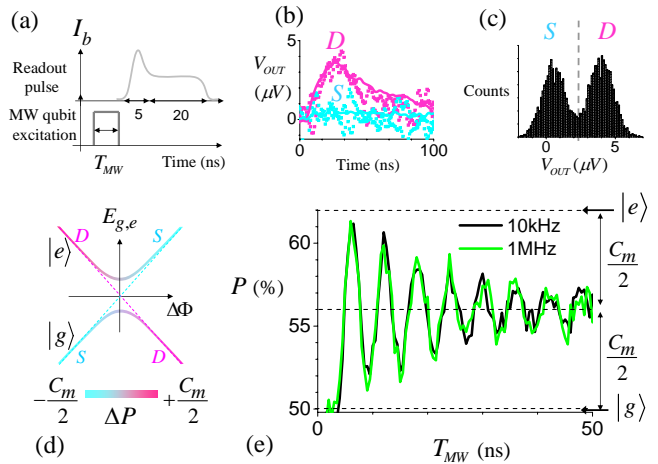}\caption{ \label{Rabi} (color on-line) (a) Current pulse for the qubit readout. (b) Output voltage. Dots correspond to real time signal, while the solid lines are average of 16 traces corresponding to the state S or D. (c) Histogram of the output voltage. (d) Energy spectrum of the flux qubit. The color scale indicates the switching probability of the detector. (e) Rabi oscillation for a measurement repetition rate of 10 kHz or 1 MHz with an identical manipulation and readout sequence at a qubit frequency of E/h=13 GHz. The three dashed lines represent P(g), P(e) and $P_m$.}
\end{figure}

The dc SQUID is operated as a DC-switching detector which is either in a purely superconducting state $S$ with zero voltage, or in a dissipative state $D$ with a finite voltage. The dc SQUID, initially in the superconducting state $S$ is biased with a current $I_b$ close to its critical current $I_c$ with a short current pulse where it might switch to the dissipative state $D$ with a probability $P$. The state of the dc SQUID is maintained using a lower bias current $I_b\approx I_c/2 $, for a time allowing to discriminate efficiently the two dc SQUID states. Figures \ref{Rabi}a and \ref{Rabi}b show the readout current pulse sent to the dc SQUID and the resulting output voltage in the case where the dc SQUID is in the state $S$ (cyan) or switches to the state $D$ (magenta). $P$ depends on the flux through the dc SQUID loop and therefore on the state of the qubit. The detector switching probabilities in the two cases when the qubit is in the ground state or in the excited state are denoted as $P(g)$ and $P(e)$. For a general qubit state with an excited state occupation probability $p(e)$ and a readout contrast $C_m \equiv P(e)-P(g)$ the switching probability is
\begin{equation}
\label{Psw-Sz}
P = P(g) + C_m p(e).
\end{equation}

Switching of the dc SQUID to its dissipative state has several severe consequences. The first trivial but important effect is that Joule heating in the shunt resistor increases the temperature of that resistor, which leads to additional noise. The dissipated energy also leads to an increased temperature of substrate and qubit. When the dc SQUID is at a nonzero voltage, the AC Josephson effect gives rise to currents at frequencies of tens to hundreds of GHz, which couple into the qubit and may cause non-adiabatic transitions and quantum leakage to higher levels. If the voltage is above the superconducting gap, generation of quasiparticles occurs in the dc SQUID junctions, deteriorating the superconducting properties. Quasiparticle generation in the qubit by transferred RF fields, and by direct diffusion when the qubit is galvanically attached, is detrimental to qubit performance. Quasiparticle recombination takes a long time and very low numbers of quasiparticles in a qubit can be significant. All these effects lead to long recuperation times for dc SQUID and qubit between successive readouts, commonly limiting the measurement repetition rate to a few tens of kHz.

To reduce the back-action of the detector, we use a very simple approach \cite{Lang,Neeley} which has never been investigated so far for non-destructive readout of superconducting qubits. By shunting the dc SQUID with a low external resistor $R$, the voltage in the dissipative state can be reduced to a value of $V \approx RI_b$ much below the superconducting gap. Therefore the generation of heat and quasiparticles is greatly suppressed. Heating is further reduced by using a fast cryogenic amplifier allowing to discriminate the two dc SQUID states efficiently with a very short readout pulse of 25 ns duration. To avoid non-adiabatic transitions of the qubit caused by high frequency oscillation of the current in the dc SQUID due to the AC-Josephson effect, the critical current of the dc SQUID is chosen such that the qubit-detector interaction is smaller than 1 GHz.

The sample is fabricated using standard e-beam lithography and shadow evaporation (Fig. \ref{sample}a and \ref{sample}b). On-chip film resistors are made of AuPd while all superconducting structures are fabricated of aluminium, with a thin aluminium oxide layer to form the Josephson junctions. The dc SQUID has two junctions of critical current $I_{c0}=80$ nA and is shunted by an interdigitated finger capacitor $C=1$ pF and a resistor $R=125$ $\Omega$. It is operated at a critical current $I_c=140$ nA, leading to $V \approx R I_c = 17$ $\mu$V. The qubit is characterised by $I_p=400$ nA and $\Delta=3.5$ GHz. It is directly coupled to the dc SQUID via a shared narrow line section with a mutual kinetic inductance $M=30$ pH. The maximum interaction between the qubit and the dc SQUID during the measurement is $2M I_p J / h=0.7$ GHz.

The dc SQUID output voltage is measured by an ultra-low noise cryogenic HEMT amplifier (Fig. \ref{sample}c), characterised by a high input impedance, $0.3$ GHz bandwidth and $\backsim 0.25$ nV/$\sqrt{\textrm{Hz}}$ noise. The amplifier is mounted on a Cu strip attached to the refrigerator still at $\sim 600$ mK. Its input is connected to the sample by two $7$ cm length superconducting wires. $R_{OUT}=3$ k$\Omega$ and $C_{OUT}\simeq 20$ pF define an overall measurement response time of 60 ns. Based on the bandwidth, the dc SQUID voltage amplitude and the amplifier noise, the two states of the dc SQUID can be discriminated to better than 97 \% confidence with a readout pulse of 25 ns total duration (Fig. \ref{Rabi}c). To gain sufficient statistics typically 25000 individual events are taken for each measurement condition. The experiments were performed in a cryogen-free dilution refrigerator, at a base temperature of 20 mK.

The resistor $R$ shunting the dc SQUID induces relaxation in the qubit. However, during the coherent manipulation of the qubit, this relaxation channel is closed by operating the dc SQUID at a bias current such that $\frac{dJ}{dIb}(I_b)=0$, where external fluctuations of $I_b$ do not couple to the dc SQUID circulating current $J$ \cite{Neeley,BertetIbstar}. For a symmetric dc SQUID, this condition is realized at $I_b=0$. During the readout, $\frac{dJ}{dIb}(I_b) \neq 0$ and the qubit relaxation rate when the dc SQUID is in its superconducting state $S$ can be estimated using Fermi's golden rule assuming a Johnson-Nyquist spectral noise density in the high-frequency limit $\hbar \omega>k_B T$
\begin{equation}
\label{T1}
\Gamma_{\downarrow} (I_b)=\frac{E}{R}\left[ \frac{\Delta}{\epsilon} \frac{M I_p }{ \hbar } \frac{dJ(I_b)}{dI_b} \right]^2.
\end{equation}
Based on (\ref{T1}), the relaxation rate due to the shunt resistor during the holding plateau is given by $1/\Gamma_{\downarrow} \simeq 500$ ns, while during the switching plateau $1/\Gamma_{\downarrow} \simeq 40$ ns. Given the total duration of the readout (25 ns) and the width of the switching plateau ($\sim 2$ ns), the relaxation during the readout should be small when the detector is in state $S$. When the dc SQUID is in its dissipative state $D$, qubit relaxation is more complex to estimate as it involves strongly non-linear effects.

We first investigate the back-action of the resistively shunted dc SQUID on the coherence of the flux qubit by performing coherent Rabi oscillations at different measurement repetition rates. As can be seen in Fig. \ref{Rabi}e, virtually no reduction of coherence is visible up to a measurement repetition rate of 1 MHz, nearly two orders of magnitude higher than previously achieved with unshunted or weakly shunted DC-switching detectors \cite{BertetReadout}. Although the Rabi coherence time of this qubit sample is relatively short, the fact that such a high measurement repetition rate is possible without any loss of readout contrast and qubit coherence shows that the back-action is very low.

\begin{figure}[t]
\includegraphics[width=3.4in]{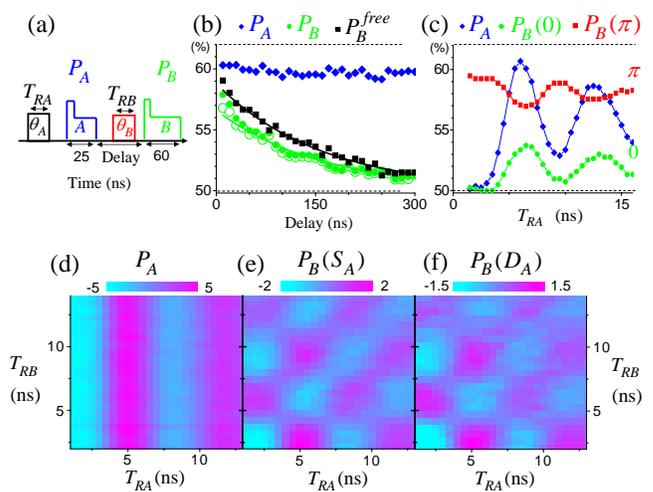}\caption{ \label{double-readout} (color on-line) (a) Qubit operation and readout sequence including two Rabi oscillations of angles $\theta_A$ and $\theta_B$, and two readout pulses A and B. The qubit operation frequency is set at E/h=13 GHz. (b) Switching probabilities $P_A, P_B$ in readout A and B as a function of the delay between the two measurements. $P^{free}_B$ is the free qubit relaxation decay measured with readout B without applying the readout pulse A. (c) $P_A, P_B$ as a function of the duration of the first Rabi pulse $T_{RA}$. $P_B(\pi)$ and $P_B(0)$ are the results readout B with or without a $\pi$ rotation between the two measurements. (d,e,f) Switching probabilities in the first readout $P_A$ and in second readout in the case that the first readout did not $P_B(S_A)$ or did switch $P_B(D_A)$ as a function of the durations $T_{RA}$ and $T_{RB}$ of both Rabi pulses.}
\end{figure}

The contrast of the readout $C_m=P(e)-P(g)$ is obtained from the Rabi oscillations in Fig. \ref{Rabi}c. As the qubit energy $E/h=13$ GHz is much higher than the thermal energy $kT/h \approx$ 1GHz, the qubit is initially in the ground state, corresponding to a detector switching probability $P(g)$. We assume that the oscillation of P as a function of the duration of the Rabi pulse $T_{MW}$ is symmetric around $P_m=(P_g+P_e)/2$, such that in the limit of a long and incoherent pulse $P=P_m$. Based on the values of $P(g)$ and $P_m$ extracted from Fig. \ref{T1}c, we deduce $C_m=12\%$.

We now turn to the analysis of the QND character of the measurement using a sequence of two successive readouts A and B and two Rabi pulses (Fig. \ref{double-readout}a). First, the switching probabilities $P_A$ and $P_B$ in the two readouts are measured as a function of the delay between the two measurements when the qubit is initially prepared in the excited state using a Rabi pulse $\theta_A=\pi$. As shown in Fig. \ref{double-readout}b, $P_A$ is constant while $P_B$ decays exponentially. To quantify the effect of readout A on the qubit, we also measure $P^{free}_B$ in the same conditions as previously except that the readout pulse A is not applied. The exponential decay of $P^{free}_B$, which is only due to the free relaxation of the qubit $T_1$=160 ns can be directly compared with the decay of $P_B$. As can be seen in Fig. \ref{double-readout}b the decay time of $P_B$ is equal to the free relaxation time (solid lines are exponential decay fitting curves). From the ratio $P_B/P^{free}_B=0.8$ we conclude that the qubit relaxation is increased by an amount of $20\ \%$ during the measurement due to the operation of the detector.

In addition to the enhancement of the qubit relaxation during the readout, we observe that the result of the first readout influences the result of the second readout for delay between the measurements shorter than 20 ns. For a delay of 10 ns, the switching probability $P_B(D_A)$ in readout B if the detector was in state $D$ during readout A is increased by about 10 \% compared to $P_B(S_A)$ if the detector was in state $S$. As this effect is not related to the qubit, it can be easily corrected for by measuring the difference $P_B(D_A)-P_B(S_A)$ with the qubit in one specific state, e.g. the ground state. In Fig. \ref{double-readout}b the open dots represent $P_B$ after applying this small correction.

Secondly, we measure $P_A$ and $P_B$ as a function of the duration $T_{RA}$ of the Rabi rotation pulse preceding readout A, with a second Rabi pulse between readout A and B, and a fixed delay time of 100 ns between the two measurements (see Fig. \ref{double-readout}a). $P_B(\pi)$ and $P_B(0)$ are obtained either with or without a $\pi$ pulse before readout B. Rabi oscillations are visible in $P_A$, $P_B(0)$ and $P_B(\pi)$ in Fig. \ref{double-readout}c. The reduced oscillation amplitude of $P_B(0)$ compared to $P_A$ is in good agreement with the free qubit relaxation during a time of 100 ns and the extra relaxation due to readout A mentioned above. The fact that the sign of the oscillation of $P_B(\pi)$ is reversed compared to $P_B(0)$ is a clear proof that the qubit state is partially preserved and can be manipulated coherently after a measurement. Furthermore, the averaged value of $P_B(\pi)$ is higher than $P_B(0)$. This is to be expected as the effect of the $\pi$ pulse is to reverse the qubit state and the qubit relaxation between the measurements is significant.

Figures \ref{double-readout}d, \ref{double-readout}e and \ref{double-readout}f show $P_A$ and the two conditional switching probabilities $P_B(S_A)$ and $P_B(D_A)$ in readout B in the cases that the result of readout A is $S$ or $D$ respectively, as a function of $T_{RA}$ and $T_{RB}$. To improve the visibility of the oscillations of $P_B(S_A)$ and $P_B(D_A)$ a constant offset is subtracted for each duration of the second Rabi pulse $T_{RB}$. As the discrimination efficiency between $|g>$ and $|e>$ in a single readout is low, the qubit state is only very weakly projected after a measurement. Given the readout contrast $C_m=12\ \%$, the correlation between readout A and B due to the qubit state projection is too small to be resolved. The two readouts are thus nearly independent. As can be seen in Fig. \ref{double-readout}e and \ref{double-readout}f, the qubit state is almost as well preserved if the detector is in state $S$ or in state $D$. This result shows that the back-action of this type of detector in the dissipative state can be reduced to a very low level.

We define the QND fidelity based on the conditional probabilities that the qubit state after the measurement is the same as before the measurement, irrespective of the measurement outcome, when the qubit is initially in $|g>$ or $|e>$, denoted as $p(g|g)$ and $p(e|e)$ respectively
\begin{equation}\label{qnd}
F_{QND}= \frac{p(g|g) + p(e|e)}{2}.
\end{equation}
This definition, which does not involve the readout contrast, is only relevant if the readout contrast is high enough compared to the available statistic. Using relation (\ref{Psw-Sz}) for the readout B, equation (\ref{qnd}) can be rewritten as $F_{QND}=\frac{1}{2}+\frac{\Delta P_B}{2 C_m}$, where $\Delta P_B = P_B(e)-P_B(g)$ is the difference of switching probability in readout B depending on the initial qubit state before readout A. From the measurement of Fig. \ref{double-readout}b for a delay between the measurement of 10 ns we find $\Delta P_B \geq 6\ \%$, without correcting for the initial qubit preparation errors but subtracting for the systematic extra switching probability due the readout A as explained previously. Assuming the contrast $C_m=12\ \%$ for readout A as well as for readout B we conclude that $F_{QND} \geq 75\ \%$, with $p(g|g)\simeq 100\ \%$ and $p(e|e)\geq 50\ \%$.

We should emphasize that this letter presents the results of the first attempt to reduce the back-action of the DC-switching readout. The readout contrast can still be largely improved while maintaining a good QND fidelity.

In conclusion, we have demonstrated that the total readout time and the back-action of the DC-switching detector can be considerably reduced, such that a single readout of a superconducting flux qubit can be completed within 25 ns with an averaged QND fidelity of $75\ \%$. Such simple, fast and low back-action detector is very attractive to study entanglement in a multiple-qubits and multiple-detectors system and opens new perspectives to investigate fundamental aspects of quantum measurement.

This work was supported by the Dutch Organization for Fundamental Research on Matter (FOM), E.U. EuroSQIP, and the NanoNed program.

\end{document}